\documentclass[10pt,twocolumn,english,aps,manuscript,showkeys]{revtex4}
\usepackage[T1]{fontenc}
\usepackage[latin9]{inputenc}
\usepackage{color}
\usepackage{amsmath}
\usepackage{amssymb}
\usepackage{graphicx}
\usepackage{wasysym}

\makeatletter
\@ifundefined{textcolor}{}
{%
 \definecolor{BLACK}{gray}{0}
 \definecolor{WHITE}{gray}{1}
 \definecolor{RED}{rgb}{1,0,0}
 \definecolor{GREEN}{rgb}{0,1,0}
 \definecolor{BLUE}{rgb}{0,0,1}
 \definecolor{CYAN}{cmyk}{1,0,0,0}
 \definecolor{MAGENTA}{cmyk}{0,1,0,0}
 \definecolor{YELLOW}{cmyk}{0,0,1,0}
}

\usepackage{mathtools,amsmath, amsthm, amssymb}

\makeatother

\usepackage{babel}
\begin{document}
\title{A conjecture on the relationship between critical residual entropy
and finite temperature pseudo-transitions of one-dimensional models}
\author{Onofre Rojas}
\email{ors@ufla.br}

\address{Departamento de Física, Universidade Federal de Lavras, 37200-900,
Lavras - MG, Brazil}
\begin{abstract}
Recently pseudo-critical temperature clues were observed in one-dimensional
spin models, such as the Ising-Heisenberg spin models, among others.
Here we report a relationship between the zero-temperature phase boundary
residual entropy (critical residual entropy) and pseudo-transition.
Usually, the residual entropy increases in the phase boundary, which
means the system becomes more degenerate at the phase boundary compared
to its adjacent states. However, this is not always the case; at zero
temperature, there are some phase boundaries where the entropy holds
the largest residual entropy of the adjacent states. Therefore, we
can propose the following conjecture: If residual entropy at zero-temperature
is a continuous function at least from the one-sided limit at a critical
point, then pseudo-transition evidence will appear at finite temperature
near the critical point. We expect that this argument would apply
to study more realistic models. Only by analyzing the residual entropy
at zero temperature, one could identify a priori whether the system
will exhibit the pseudo-transition at finite temperature. To strengthen
our conjecture, we use two examples of Ising-Heisenberg models, which
exhibit pseudo-transition behavior: one frustrated coupled tetrahedral
chain and another unfrustrated diamond chain.
\end{abstract}
\maketitle

\section{Introduction}

In 1950, van Hove\citep{Hove} proposed a theorem to rigorously verify
the absence of phase transition with a short-range interaction for
a uniform one-dimensional system. This theorem is valid under the
following conditions: (i) Homogeneity, excluding automatically inhomogeneous
system, i.e., disordered or periodic. (ii) The Hamiltonian does not
include particles position terms, e.g., external fields. (iii) Hard-core
particles, this means the theorem cannot be applied to point-like
or soft particles. The theorem has been proved writing the partition
function, by using the transfer-matrix technique and reducing the
problem to find the largest eigenvalue, which implies that the free
energy is an analytic function. In fact, the theorem proves that the
one-dimensional models with short-range coupling do not exhibit any
phase transitions. Mermin and Wagner\citep{Mermin} also rigorously
proved the absence of ferromagnetism or antiferromagnetism in one
and two dimensional isotropic Heisenberg model. Recently, Cuesta and
Sanchez\citep{cuesta} proposed a more general non-existence theorem
for phase transition at finite temperature. Mainly they included an
external field and considered point-like particles, which broadens
the non-existence theorem. But it is not yet a fully general theorem,
e.g., no mixed particle chains and more general external fields were
included.

Despite of that, there are some one-dimensional models with a short-range
coupling that exhibit a first-order phase transition at finite temperature.
The Kittel model (also known as the zipper model)\citep{kittel},
is a typical simple model with a finite size transfer-matrix.  Note
that the constraint on zipper corresponds to an infinite potential,
and this condition leads to a non-analytic free energy. Consequently,
the system exhibits a first-order phase transition. Other model is
that considered by Chui-Weeks model\citep{chui}, with a typical set
of models called solid-on-solid for surface growth. While in this
case the transfer-matrix dimension is infinite, but it can still be
solved exactly. Furthermore, imposing the impenetrable condition to
subtract, the model shows the existence of phase-transition. Dauxois-Peyrard
model\citep{dauxois}, is another model with infinite transfer-matrix
dimension, which can be solved numerically. More recently Sarkanych
et al.\citep{sarkanych} also proposed a one-dimensional Potts model
with invisible states and short-range coupling. The term invisible
essentially refers to an additional energy degeneracy, which contributes
to the entropy, but not the interaction energy. So, these invisible
states are responsible for generating the first-order phase transition.
In a nutshell, all these models break the Perron-Frobenius theorem\citep{ninio},
because the free energy becomes non-analytical at the phase transition
temperature, or equivalently some elements of transfer-matrix become
null (which corresponds to an infinite energy).

On the other hand, the term \textquotedbl pseudo-transition\textquotedbl{}
was introduced by Timonin\citep{Timonin} in 2011 while studying the
spin ice in a field, and refers to a sudden change in the first derivative
of free energy, whereas a strong vigorous peak appears in the second
derivative of free energy, although there are no discontinuity or
divergence, respectively. Later this definition was adopted for our
group\citep{Isaac,Isaac 2,pseudo} because we found the same kind
of property. The pseudo-transition does not violate the Perron-Frobenius
theorem\citep{ninio}, since the free energy is always analytic.  The
anomalous behavior occurs only because off diagonal elements of the
transfer-matrix become a tiny amount compared to other elements.

Recent investigations revealed a number of decorated one-dimensional
models, particularly the Ising and Heisenberg models with a variety
of structures. Such as the Ising-Heisenberg diamond chain\citep{torrico,torrico2}
and even Ising diamond chain\citep{Strecka-ising}. One-dimensional
double-tetrahedral model, where the nodal site is assembled by localized
Ising spin, and alternating with a pair of delocalized mobile electrons
within a triangular plaquette\citep{Galisova}.  Ladder model with
alternating Ising-Heisenberg coupling\citep{on-strk}. As well as
the triangular tube model with Ising-Heisenberg coupling \citep{strk-cav}.
In all aforementioned models, pseudo-transition clues were observed.
The first derivative of free energy, such as entropy, internal energy,
and magnetization, shows a steep but still continuous change as the
temperature varies, which is quite similar to the first-order phase
transition behavior.  While the second-order derivative of free energy,
like the specific heat and magnetic susceptibility, resembles a typical
second-order phase transition behavior at finite temperature. Therefore,
this peculiar behavior drew attention to a more careful study, as
considered in reference \citep{pseudo}. Lately, in reference\citep{Isaac}
has been made an additional discussion on this property and detailed
study of the correlation function for arbitrarily distant spins around
the pseudo-transition.

The article is organized as follows: In sec. 2 we present the free
energy for spin-1/2 like one-dimensional model and we analyze the
low temperatures behavior. In Sec.3 we present the residual entropy
at a critical point, and its connection with the pseudo-transition
at finite temperature. In Sec.4 we apply it to a frustrated Ising-Heisenberg
coupled tetrahedral chain. Analogously, in Sec. 5, we also apply it
to an unfrustrated Ising-Heisenberg diamond chain. Finally, Sec.6
provides our conclusions and perspectives.

\section{Free energy}

The models commented above can be seen as decorated models\citep{torrico,torrico2,Galisova,on-strk,strk-cav},
which can be mapped to a simple spin-1/2 Ising-like model\citep{Dec-trnsf}.
Like models considered here {[}see for instance in eqs. \eqref{eq:H-tetra}
and \eqref{eq:Hamt}{]}, can be mapped\citep{Dec-trnsf} into an effective
Hamiltonian like 
\begin{equation}
H_{{\rm eff}}=-\sum_{i=1}^{N}\left[K_{0}+Ks_{i}s_{i+1}+\tfrac{1}{2}B(s_{i}+s_{i+1})\right],\label{eq:Ham-1}
\end{equation}
 where $K_{0}$, $K$ and $B$ are effective parameters, which may
depend on the temperature and original Hamiltonian parameters (for
details see for instance ref. \citep{torrico,torrico2,Galisova,on-strk,strk-cav}),
assuming the chain contains $N$ unit cells. Therefore, the Hamiltonian
\eqref{eq:Ham-1} transfer-matrix, can be expressed as $\mathbf{V}=\left[\begin{array}{cc}
w_{1} & w_{0}\\
w_{0} & w_{-1}
\end{array}\right]$, like discussed in reference \citep{pseudo}. The transfer-matrix
entries $w_{n}$ (Boltzmann factor) become 
\begin{equation}
w_{n}=\sum_{k=0}g_{n,k}{\rm e}^{-\beta\varepsilon_{n,k}},\label{eq:G-w}
\end{equation}
with $n=\{-1,0,1\}$ (denoted by sectors -1, 0, 1). Here $\varepsilon_{n,k}$
represent the energy spectra $k=\{0,1,\ldots\}$ for each sector defined
above (not for whole system), and $g_{n,k}$ denotes the degeneracy
for each energy level, where we assume $g_{n,k}=\{1,2,3,\dots\}$.
Whereas $\beta=1/k_{B}T$, with $k_{B}$ being the Boltzmann constant
and $T$ is the absolute temperature.

Then the transfer-matrix eigenvalues are provided by
\begin{equation}
\lambda_{\pm}=\tfrac{1}{2}\Bigl(w_{1}+w_{-1}\pm\sqrt{(w_{1}-w_{-1})^{2}+4w_{0}^{2}}\Bigr).\label{eq:L12}
\end{equation}
Assuming the chain has a periodic boundary condition, so the partition
function becomes $\mathcal{Z}_{N}=\lambda_{+}^{N}+\lambda_{-}^{N}$.
Consequently, the free energy in the thermodynamic limit ($N\rightarrow\infty$)
results in
\begin{equation}
f=-\tfrac{1}{\beta}\ln\!\left[\tfrac{1}{2}\Bigl(w_{1}+w_{-1}+\sqrt{(w_{1}-w_{-1})^{2}+4w_{0}^{2}}\Bigr)\right]\!.\label{eq:free-energ}
\end{equation}

\subsection{Low temperature free energy when $w_{0}\rightarrow0$}

As earlier discussed in the literature\citep{pseudo}, if we consider
$w_{0}=0$, free energy \eqref{eq:free-energ} reduces to 
\begin{equation}
f=-\tfrac{1}{\beta}\ln\left[\max\left(w_{1},w_{-1}\right)\right].\label{eq:free-dsc}
\end{equation}

This result could mean the presence of a genuine phase transition
at finite temperature, because \eqref{eq:free-dsc} becomes a non-analytic
function when $w_{1}=w_{-1}$. Obviously, this cannot happen in this
limit, because $w_{0}$ is small enough but not null.

Now, in general, the energies $\varepsilon_{_{n,k}}$, as defined
above, should depend on some parameters, here we simply denote by
$x$, e.g., magnetic field or some other parameters. Therefore, let
us conveniently define the following quantities: $\bar{\varepsilon}(x)=\frac{\varepsilon_{1,0}(x)+\varepsilon_{-1,0}(x)}{2}$
being the average between lowest energies of both sectors; another
quantity we define, is the difference between lowest energies in different
sectors $\epsilon(x)=\varepsilon_{1,0}(x)-\varepsilon_{-1,0}(x)$;
while $\delta_{1}(x)=\varepsilon_{1,1}(x)-\varepsilon_{1,0}(x)$ and
$\delta_{-1}(x)=\varepsilon_{-1,1}(x)-\varepsilon_{-1,0}(x)$ are
energy differences within same sector.

Next, let us consider the Boltzmann factor with good accuracy in the
low temperatures region, including only the ground state energy and
the lowest excited state energy for sectors -1 and 1. Thus we have

\begin{alignat}{1}
w_{1}(x,T)= & g_{1,0}{\rm e}^{-\beta\varepsilon_{1,0}(x)}+g_{1,1}{\rm e}^{-\beta\varepsilon_{1,1}(x)}\nonumber \\
= & {\rm e}^{-\beta\bar{\varepsilon}(x)}\eta_{1}(x,T),\label{eq:w1xT}\\
w_{-1}(x,T)= & g_{-1,0}{\rm e}^{-\beta\varepsilon_{-1,0}(x)}+g_{-1,1}{\rm e}^{-\beta\varepsilon_{-1,1}(x)}\nonumber \\
= & {\rm e}^{-\beta\bar{\varepsilon}(x)}\eta_{-1}(x,T),\label{eq:w-1xT}
\end{alignat}
where we define

\begin{equation}
\eta_{\nu}(x,T)=g_{\nu,0}{\rm e}^{-\nu\frac{\epsilon(x)}{2T}}\left[1+\tfrac{g_{\nu,1}}{g_{\nu,0}}{\rm e}^{-\frac{\delta_{\nu}(x)}{T}}\right],\label{eq:eta-nu}
\end{equation}
with $\nu=\pm1$.

The free energy \eqref{eq:free-dsc} can be rewritten in the neighboring
of quantum phase transition as a function of temperature $T$ and
parameter $x$, 
\begin{equation}
f(x,T)=\bar{\varepsilon}(x)-T\ln\left\{ \max\left[\eta_{1}(x,T),\eta_{-1}(x,T)\right]\right\} .\label{eq:low-free-enrg}
\end{equation}

From now on, we will focus on the thermodynamic properties when $w_{1}$
and $w_{-1}$ are competing terms and under the condition $w_{0}\rightarrow0$.

\section{Residual entropy and pseudo-transition}

In this section, we will discuss residual entropy at zero temperature,
and particularly we stress residual entropy around the ground state
phase transition. This phase transition will occur by varying some
control parameter like $x$, at a critical point $x_{c}$. The entropy,
at this point, will simply be called critical residual entropy (CRE).
At lats, we observe that CRE has a relationship with the finite temperature
pseudo-transition\citep{rd-entr-O}.

\subsection{On the continuity of residual entropy}

At zero temperature, the critical residual entropy (CRE) occurs at
the phase boundary of two ground states. For this purpose, we can
use the free energy given in \eqref{eq:free-energ} when $T\rightarrow0$.
Hence, varying a given parameter $x$, we can get the residual entropy
at zero temperature. Considering that the zero temperature phase transition
arises at $x_{c}$, then adjacent states coexist, where we denote
the critical energy $\varepsilon_{c}$ and corresponding critical
degeneracy defined by $G_{c}$.

Now let us continue assuming the condition of the previous section,
that is $w_{0}\rightarrow0$, while $w_{1}$ and $w_{-1}$ are competing
terms. Thus, the lowest energies for each sectors are $\varepsilon_{_{1,0}}(x)$
and $\varepsilon_{_{-1,0}}(x)$, in the interface we have $\varepsilon_{_{1,0}}(x_{c})=\varepsilon_{_{-1,0}}(x_{c})=\varepsilon_{c}$.
For a particular case when $x\rightarrow x_{c}$, we have the following
limits in the expression \eqref{eq:eta-nu}: $\ensuremath{\lim\limits _{x\rightarrow x_{c}}\frac{\delta_{\pm1}(x)}{T}\rightarrow+\infty}$,
$\ensuremath{\lim\limits _{x\rightarrow x_{c}}\epsilon(x)=0}$ and
$\ensuremath{\lim\limits _{x\rightarrow x_{c}}\varepsilon(x)=\varepsilon_{c}}$.
So the free energy \eqref{eq:low-free-enrg} reduces to 
\begin{equation}
f(x_{c},T)=\varepsilon_{c}-T\ln\left[\max\left(g_{1,0},g_{-1,0}\right)\right].
\end{equation}
Afterward, we can obtain the corresponding CRE at zero temperature
\begin{equation}
\mathcal{S}_{c}=\ln\left[\max\left(g_{1,0},g_{-1,0}\right)\right].\label{eq:Sg-psd}
\end{equation}
Through this article we will consider the entropy in units of $k_{B}$.
Moreover, the critical degeneracy per unit cell results in $G_{c}=\max\left(g_{1,0},g_{-1,0}\right)$. 

It is worth mentioning, according to the third law of thermodynamics
or often referred to as the Nernst's postulate. At zero temperature,
entropy leads to a constant and must be independent of any parameter
(such as $x$), so residual entropy is determined only by its ground
state energy degeneracy.

\begin{figure}[h]
\centering{}\includegraphics[scale=0.9]{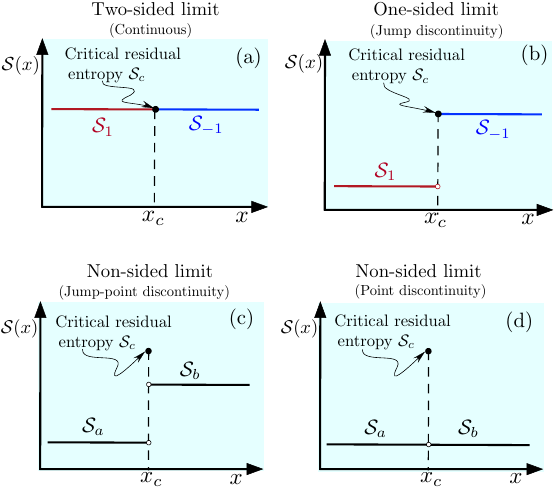}\caption{\label{fig:entropy-T0}(a) Typical continuous residual entropy at
zero temperature as a function of parameter $x$, assuming the system
has two phases; (b) Continuous residual entropy at zero temperature
from the one-sided limit at $x_{c}$; (c-d) Typical discontinuous
residual entropy from the non-sided limit at $x_{c}$, the CRE is
strictly larger than neighboring residual entropy.}
\end{figure}

Below we present an accurate mathematical formulation\citep{bk-stewart}
of the residual entropy around the critical point found in \eqref{eq:Sg-psd}.

When $g_{1,0}=g_{-1,0}$, residual entropy is illustrated schematically
in fig.\ref{fig:entropy-T0}a. Therefore, entropy as a function of
$x$ at zero temperature, has left and right limits, 
\begin{equation}
\lim_{x\rightarrow x_{c}^{-}}\mathcal{S}(x)=\lim_{x\rightarrow x_{c}^{+}}\mathcal{S}(x)=\mathcal{S}(x_{c})=\mathcal{S}_{c},\label{eq:Sc-contns}
\end{equation}
and both limits are identical, then we say the residual entropy is
continuous at $x_{c}$. However, it is worth mentioning that we are
considering two different adjacent phases, which physically correspond
to two different states, with identical residual entropies \eqref{eq:Sc-contns}.
We will see this case later when we apply it to a peculiar unfrustrated
model.

When $g_{1,0}\ne g_{-1,0}$, residual entropy is illustrated schematically
in fig.\ref{fig:entropy-T0}b. Assuming that the degeneracies of the
adjacent phases satisfy $g_{-1,0}<g_{1,0}$, then the left and right
limit of the residual entropy at $x_{c}$ becomes, 
\begin{equation}
\lim_{x\rightarrow x_{c}^{-}}\mathcal{S}(x)<\lim_{x\rightarrow x_{c}^{+}}\mathcal{S}(x)=\mathcal{S}(x_{c})=\mathcal{S}_{c}.\label{eq:Sc-dscntns}
\end{equation}
In this case, the residual entropy is continuous from the right-sided
limit, but discontinuous from the left-sided limit\citep{bk-stewart}.
Consequently, we can say that the residual entropy is continuous from
the one-sided limit at $x_{c}$. Opposite condition $g_{-1,0}>g_{1,0}$,
can be obtained in a similar way.

On the other hand, in Appendix \ref{Appdx-A}, we present some detailed
results for discontinuous residual entropy, where we considered a
number of combinations among other sectors. In all those cases, critical
degeneracy is strictly larger than the surrounding degeneracy $G_{c}>\max\left(g_{1,0},g_{-1,0}\right)$. 

Therefore, residual entropy must satisfy the following relation 
\begin{equation}
\lim_{x\rightarrow x_{c}^{\pm}}\mathcal{S}(x)<\mathcal{S}(x_{c})=\mathcal{S}_{c},\label{eq:S-sntd}
\end{equation}
which is discontinuous at $x_{c}$. Because there is no left or right
sided limit\citep{bk-stewart}, so it has a jump-point (fig.\ref{fig:entropy-T0}c)
and point (fig.\ref{fig:entropy-T0}d) discontinuity at $x_{c}$.
The fig.\ref{fig:entropy-T0}d cannot be confused with a \textquotedblleft removable\textquotedblright{}
point discontinuity, because the CRE is strictly larger than the neighboring
residual entropy. Note that $\mathcal{S}_{a}$ and $\mathcal{S}_{b}$
in fig.\ref{fig:entropy-T0}c-d can represent either $\mathcal{S}_{1}$
or $\mathcal{S}_{-1}$. 

\subsection{Pseudo-critical temperature}

As discussed previously in ref. \citep{Isaac,pseudo}, the pseudo-critical
temperature can be found using the following relation
\begin{equation}
w_{1}(x_{p},T_{p})=w_{-1}(x_{p},T_{p}),\label{eq:w1-w2}
\end{equation}
where $x_{p}$ and $T_{p}$ correspond to the Hamiltonian parameters
at the pseudo-transition point.

As a first approximation, we can consider only the lowest energy for
each sector $n=1$ and $-1$. Thereby, the Boltzmann factors become
$w_{1}=g_{1,0}{\rm e}^{-\beta\varepsilon_{1,0}}$ and $w_{-1}=g_{-1,0}{\rm e}^{-\beta\varepsilon_{-1,0}}$.

Besides assuming $\epsilon_{p}=\epsilon(x_{p})$ in \eqref{eq:w1-w2},
we obtain the following relation, 
\begin{equation}
{\rm e}^{-\frac{\epsilon_{p}}{T_{p}}}=\frac{g_{-1,0}}{g_{1,0}}.
\end{equation}
 Then we get a simple expression for the pseudo-critical temperature
\begin{equation}
T_{p}=\frac{\epsilon_{p}}{\ln\left(\frac{g_{1,0}}{g_{-1,0}}\right)}=\frac{\varepsilon_{1,0}(x_{p})-\varepsilon_{-1,0}(x_{p})}{\ln\left(\frac{g_{1,0}}{g_{-1,0}}\right)}.\label{eq:gen-eq-ws-1}
\end{equation}

Note that result \eqref{eq:gen-eq-ws-1} has already been discussed
in reference \citep{Galisova,on-strk,strk-cav}. It is also worth
mentioning that the critical temperature for the Kittel model\citep{kittel}
has a quite similar expression.

However, eq.\eqref{eq:gen-eq-ws-1} fails in the case of $g_{1,0}=g_{-1,0}$
because $T_{p}$ turns undefined. To improve \eqref{eq:gen-eq-ws-1},
we need to include the lowest excited energy in at least one sector.
Hence, using eq.\eqref{eq:eta-nu} in eq.\eqref{eq:w1-w2}, we obtain

\begin{equation}
\eta_{1}(x_{p},T_{p})=\eta_{-1}(x_{p},T_{p}).
\end{equation}
Even more explicitly, we can write in terms of a transcendental equation
as follows
\begin{alignat}{1}
{\rm e}^{-\frac{\epsilon_{p}}{T_{p}}} & =\frac{g_{-1,0}+g_{-1,1}{\rm e}^{-\frac{\delta_{-1,p}}{T_{p}}}}{g_{1,0}+g_{1,1}{\rm e}^{-\frac{\delta_{1,p}}{T_{p}}}},\nonumber \\
 & =\left(\frac{g_{-1,0}}{g_{1,0}}\right)\frac{1+\frac{g_{-1,1}}{g_{-1,0}}{\rm e}^{-\frac{\delta_{-1,p}}{T_{p}}}}{1+\frac{g_{1,1}}{g_{1,0}}{\rm e}^{-\frac{\delta_{1,p}}{T_{p}}}},\label{eq:gen-eq-ws}
\end{alignat}
where $\epsilon_{p}=\epsilon(x_{p})$, $\delta_{1,p}=\delta_{1}(x_{p})$
and $\delta_{-1,p}=\delta_{-1}(x_{p})$ are the aforementioned energy
differences.

\begin{figure}[h]
\centering{}\includegraphics[scale=0.55]{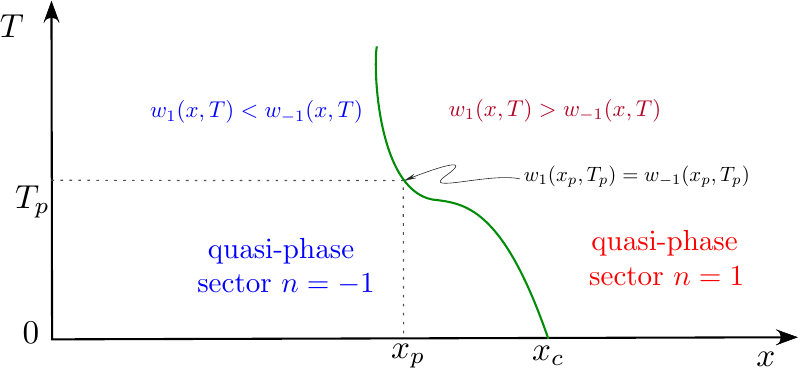}\caption{\label{fig:QP-dmg}Typical quasi-phase\citep{Timonin} diagram $x$
against $T$. Real phase transition occurs only at zero temperature
$(x_{c},0)$. For finite temperature ($T>0$) arises a pseudo-transition
at $(x_{p},T_{p})$.}
\end{figure}

Usually, when the degeneracies in the ground states are unequal, it
is enough to use \eqref{eq:gen-eq-ws-1}. Nevertheless, when the ground
state degeneracies are identical, we need to use eq.\eqref{eq:gen-eq-ws}.
Surely, we can also use the full expression given by \eqref{eq:w1-w2},
which we can readily solve by numerical computation.

In fig.\ref{fig:QP-dmg} we schematically illustrate a typical pseudo-transition
curve given by eq.\eqref{eq:w1-w2}. Here, we remark that a true phase
transition occurs only at zero temperature and for a given critical
point $x_{c}$. 

In ref.\citep{karlova} a similar approach was considered when analyzing
the maximum peak for the specific heat, and the peak height is related
to the degeneracy of the ground state energy. 

In summary, the most interesting result from the above study leads
to the following conjecture: If zero-temperature residual entropy
is continuous at a critical point at least from the one-sided limit,
then we can observe vestiges of finite temperature pseudo-transition
near the critical point.

Residual entropy, as illustrated in fig.\ref{fig:entropy-T0}a is
a differentiable two-sided function\citep{bk-stewart} at a critical
point \eqref{eq:Sc-contns}. Due to the differentiability of residual
entropy, as soon as temperature increases, entropy should increase
smoothly around $x_{c}$, bringing relevant information from adjacent
ground state phases without significant disturbances.

Similarly, residual entropy described by fig.\ref{fig:entropy-T0}b
is a one-sided differentiable function at a critical point\eqref{eq:Sc-dscntns}.
Therefore, as in the previous case, the one-sided differentiability
takes into account relevant information about the adjacent ground
states, which should persist without significant disturbance as temperature
increases.

In contrast, the discontinuous entropy given by eq.\eqref{eq:S-sntd}
is a non-differentiable function at a critical point. Therefore, we
observe as soon as the temperature increases, the adjacent phases
near the critical point stretched around critical residual entropy
(single point), destroying any evidence of zero-temperature phase
transition as temperature increases. So we do not observe a pseudo-transition
at finite temperature. 

\section{Ising-Heisenberg coupled tetrahedral chain}

\begin{figure}[h]
\centering{}\includegraphics[scale=0.8]{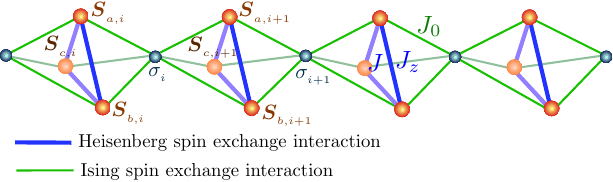}\caption{\label{fig:Sch-trd-ch}Schematic representation of Ising-Heisenberg
coupled tetrahedral chain. Small spheres ($\sigma_{i}$) corresponds
to Ising spins, and large spheres ($\boldsymbol{S}_{a(b),i}$) correspond
to Heisenberg spins.}
\end{figure}

Earlier, in the reference \citep{mambrini,roj-alc}, the Heisenberg
version of the coupled tetrahedral chain was investigated. While Galisova
and Strecka\citep{Galisova,galisova17}, considered delocalized electrons
and Ising spin in the tetrahedral chain. Later, in the reference \citep{vadim-1,Vadim-2}
the Ising-Heisenberg version of the model was introduced. Similar
model with higher spin was also analyzed more recently in reference
\citep{tetra-spin1}. 

Here we explore a slightly different model (see fig.\ref{fig:Sch-trd-ch}),
which exhibits a pseudo-transition property that has not yet been
discussed. So, we present the Hamiltonian of the model as follows
\begin{alignat}{1}
H= & -\sum_{i=1}^{N}\left\{ J(\boldsymbol{S}_{a,i},\boldsymbol{S}_{b,i})_{z}+J(\boldsymbol{S}_{b,i},\boldsymbol{S}_{c,i})_{z}\right.\nonumber \\
 & +J(\boldsymbol{S}_{c,i},\boldsymbol{S}_{a,i})_{z}+\tfrac{h}{2}\left(\sigma_{i}+\sigma_{i+1}\right)\nonumber \\
 & \left.+\left(S_{a,i}^{z}+S_{b,i}^{z}+S_{c,i}^{z}\right)\left[h_{z}+J_{0}(\sigma_{i}+\sigma_{i+1})\right]\right\} ,\label{eq:H-tetra}
\end{alignat}
where $J(\boldsymbol{S}_{a,i},\boldsymbol{S}_{b,i})_{z}=JS_{a,i}^{x}S_{b,i}^{x}+JS_{a,i}^{y}S_{b,i}^{y}+J_{z}S_{a,i}^{z}S_{b,i}^{z}$,
with $S_{a,i}^{\alpha}$ denoting the Heisenberg spin-1/2, and $\alpha=\{x,y,z\}$,
while $\sigma_{i}$ denotes the Ising spin ($\sigma_{i}=\pm\frac{1}{2}$).
In a similar way we define for sites $b$ and $c$ in \eqref{eq:H-tetra}.
While $J$ describes the exchange interaction of Heisenberg spin in
$xy$-anisotropy, similarly $J_{z}$ stands for $z$-anisotropy exchange
interaction, and by $J_{0}$ we denote the Ising-Heisenberg exchange
interaction. Whereas $h$ and $h_{z}$ correspond to the magnetic
field acting on the Ising and Heisenberg spins. 

Within the triangle structure, Heisenberg spins must compose the $8\times8$
dimension operator. But this operator we can express as block matrices,
one quadruplet and two doublet states, which can be readily diagonalized.
The quadruplet have two eigenvalues: the first one is
\begin{alignat*}{1}
e_{_{\frac{3}{2},\frac{3}{2}}}= & -\frac{3J_{z}}{4},
\end{alignat*}
with corresponding eigenvectors
\begin{equation}
\left|\tfrac{3}{2},+\tfrac{3}{2}\right\rangle =\left|\substack{+\\
+\\
+
}
\right\rangle \;\text{and}\;\left|\tfrac{3}{2},-\tfrac{3}{2}\right\rangle =\left|\substack{-\\
-\\
-
}
\right\rangle ;
\end{equation}
the second eigenvalue is 
\begin{alignat*}{1}
e_{_{\frac{3}{2},\frac{1}{2}}}= & -J+\frac{J_{z}}{4},
\end{alignat*}
whose eigenvectors are given by
\begin{alignat}{1}
\left|\tfrac{3}{2},+\tfrac{1}{2}\right\rangle = & \tfrac{1}{\sqrt{3}}\left(\left|\substack{+\\
+\\
-
}
\right\rangle +\left|\substack{+\\
-\\
+
}
\right\rangle +\left|\substack{-\\
+\\
+
}
\right\rangle \right),\nonumber \\
\left|\tfrac{3}{2},-\tfrac{1}{2}\right\rangle = & \tfrac{1}{\sqrt{3}}\left(\left|\substack{-\\
-\\
+
}
\right\rangle +\left|\substack{-\\
+\\
-
}
\right\rangle +\left|\substack{+\\
-\\
-
}
\right\rangle \right);
\end{alignat}
both energy levels are twofold degenerate.

While the eigenvalue of the doublet pair is
\begin{alignat*}{1}
e_{_{\frac{1}{2},\frac{1}{2}}}= & \frac{J}{2}+\frac{J_{z}}{4},
\end{alignat*}
this energy level is fourfold degenerate.

The first doublet eigenstates become

\begin{equation}
\begin{cases}
\left|\tfrac{1}{2},+\tfrac{1}{2}\right\rangle = & \frac{1}{\sqrt{6}}\left(\left|\substack{+\\
+\\
-
}
\right\rangle -2\left|\substack{+\\
-\\
+
}
\right\rangle +\left|\substack{-\\
+\\
+
}
\right\rangle \right),\\
\left|\tfrac{1}{2},+\tfrac{1}{2}\right\rangle = & \frac{1}{\sqrt{2}}\left(\left|\substack{-\\
+\\
+
}
\right\rangle -\left|\substack{+\\
+\\
-
}
\right\rangle \right),
\end{cases}
\end{equation}
while the second doublet eigenstates are

\begin{equation}
\begin{cases}
\left|\tfrac{1}{2},-\tfrac{1}{2}\right\rangle = & \frac{1}{\sqrt{6}}\left(\left|\substack{-\\
-\\
+
}
\right\rangle -2\left|\substack{-\\
+\\
-
}
\right\rangle +\left|\substack{+\\
-\\
-
}
\right\rangle \right),\\
\left|\tfrac{1}{2},-\tfrac{1}{2}\right\rangle = & \frac{1}{\sqrt{2}}\left(\left|\substack{-\\
-\\
+
}
\right\rangle -\left|\substack{+\\
-\\
-
}
\right\rangle \right).
\end{cases}
\end{equation}

\subsection{Zero temperature phase diagram}

Now using the eigenvalues found above, we can express the energy levels
and corresponding degeneracies, 

\begin{alignat}{2}
\varepsilon_{n,0}= & \frac{1}{2}\left(J_{0}-h\right)n-J+\frac{J_{z}}{4}+\frac{h_{z}}{2}, & g_{n,0}= & 1,\label{eq:e-Trd-0}\\
\varepsilon_{n,1}= & \frac{1}{2}\left(3J_{0}-h\right)n-\frac{3J_{z}}{4}+\frac{3h_{z}}{2}, & g_{n,1}= & 1,\\
\varepsilon_{n,2}= & {\normalcolor {\normalcolor {\color{red}{\normalcolor \frac{1}{2}\left(J_{0}-h\right)n+\frac{J}{2}+\frac{J_{z}}{4}+\frac{h_{z}}{2}}},}} & g_{n,2}= & 2,\\
\varepsilon_{n,3}= & {\color{blue}{\normalcolor -\frac{1}{2}\left(J_{0}+h\right)n+\frac{J}{2}+\frac{J_{z}}{4}-\frac{h_{z}}{2}}},\quad & g_{n,3}= & 2,\\
\varepsilon_{n,4}= & {\color{green}{\normalcolor -\frac{1}{2}\left(3J_{0}+h\right)n-\frac{3J_{z}}{4}-\frac{3h_{z}}{2}}},\quad & g_{n,4}= & 1,\\
\varepsilon_{n,5}= & -\frac{1}{2}\left(J_{0}+h\right)n-J+\frac{J_{z}}{4}-\frac{h_{z}}{2},\quad & g_{n,5}= & 1,\label{eq:e-Trd-5}
\end{alignat}
 where $n=\{-1,0,1\}$. To analyze the phase diagram at zero temperature,
we explore some relevant states below. 

\begin{figure}[h]
\begin{centering}
\includegraphics[scale=0.44]{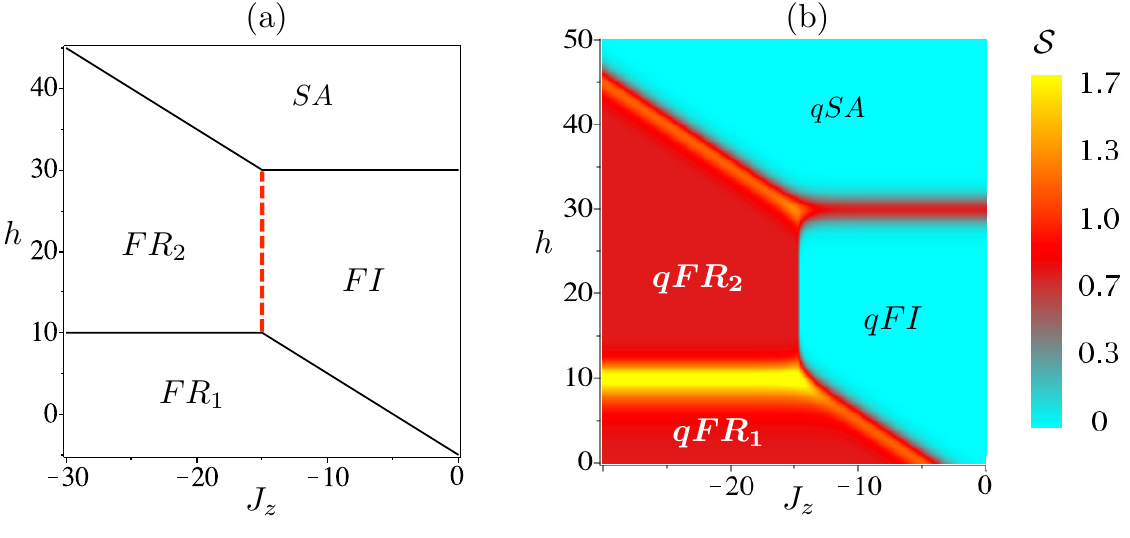}
\par\end{centering}
\caption{\label{fig:(a)-Zero-temperature}(a) Zero temperature the phase diagram
for Ising-Heisenberg coupled tetrahedral chain in the $J_{z}-h$ plane,
assuming the fixed parameters $J=-10$, $J_{0}=-10$ and $h_{z}=h$;
(b) Entropy density plot for temperature $T=0.6$, assuming the same
set of parameters considered in (a).}
\end{figure}

First, we report the ground state energy of the saturated phase (SA),
which reads as
\begin{alignat*}{1}
E_{SA}=\varepsilon_{1,4}= & -\frac{1}{2}\left(3J_{0}+h\right)-\frac{3J_{z}}{4}-\frac{3h_{z}}{2}.
\end{alignat*}
 While its ground state becomes 
\begin{equation}
|SA\rangle=\prod_{i=1}^{N}\left|\tfrac{3}{2},+\tfrac{3}{2}\right\rangle _{i}|\uparrow\rangle_{i},\label{Trt-SA}
\end{equation}
whose Ising spin magnetization per unit cell is $m_{I}=\frac{1}{2}$,
and Heisenberg spin magnetization per unit cell is $m_{H}=\frac{3}{2}$,
with total spin magnetization per unit cell $m_{t}=m_{I}+m_{H}=2$.

Second, the ground state energy for ferrimagnetic (FI) phase can be
expressed as
\begin{alignat*}{1}
E_{FI}=\varepsilon_{-1,4}= & \frac{1}{2}\left(3J_{0}+h\right)-\frac{3J_{z}}{4}-\frac{3h_{z}}{2},
\end{alignat*}
and its ground state and magnetizations turns in,
\begin{equation}
|FI\rangle=\prod_{i=1}^{N}\left|\tfrac{3}{2},+\tfrac{3}{2}\right\rangle _{i}|\downarrow\rangle_{i},\label{Trt-FI}
\end{equation}
with $m_{I}=-\frac{1}{2}$, $m_{H}=\frac{3}{2}$ and $m_{t}=1$.

The third phase we consider is a frustrated phase, whose ground state
energy is given by
\begin{alignat*}{1}
E_{_{FR_{1}}}=\varepsilon_{1,2}= & \frac{1}{2}\left(J_{0}-h\right)+\frac{J}{2}+\frac{J_{z}}{4}+\frac{h_{z}}{2},
\end{alignat*}
 with corresponding ground state
\begin{equation}
|FR_{1}\rangle=\prod_{i=1}^{N}\left|\tfrac{1}{2},-\tfrac{1}{2}\right\rangle _{i}|\uparrow\rangle_{i},\label{Trt-FR1}
\end{equation}
whose magnetizations are $m_{I}=\frac{1}{2}$, $m_{H}=-\frac{1}{2}$
and $m_{t}=0$. 

Fourth, one additional frustrated ground state energy is considered
\begin{alignat*}{1}
E_{_{FR_{2}}}=\varepsilon_{1,3}= & -\frac{1}{2}\left(J_{0}+h\right)+\frac{J}{2}+\frac{J_{z}}{4}-\frac{h_{z}}{2},
\end{alignat*}
 and its respective ground state is represented by
\begin{equation}
|FR_{2}\rangle=\prod_{i=1}^{N}\left|\tfrac{1}{2},+\tfrac{1}{2}\right\rangle _{i}|\uparrow\rangle_{i},\label{Trt-FR2}
\end{equation}
with magnetizations $m_{I}=\frac{1}{2}$, $m_{H}=\frac{1}{2}$ and
$m_{t}=1$.

In fig.\ref{fig:(a)-Zero-temperature}a, the phase diagram is shown
at zero temperature, where ground state is described in each region. 
The phase boundary between $FR_{1}$ and $FR_{2}$ is given by $h=10$,
whose phase boundary degeneracy is composed by $g_{1,0}^{c}=2$, $g_{1,1}^{c}=2$,
$g_{-1,0}^{c}=2$ and $g_{0,0}^{c}=2$, then using the eq.\eqref{eq:Sg3},
the CRE becomes ${\cal S}_{c}=\ln(3+\sqrt{5})$. The straight line
describing the interface between $FR_{1}$ and $FI$ is given by $h=-J_{z}+0.5$.
Hence, the CRE can be obtained using eq.\eqref{eq:S_01}, so we have
$\mathcal{S}_{c}=\ln(3)$, because $g_{1,0}^{c}=2$ and $g_{-1,1}^{c}=1$.
In a similar way, the boundary between $FI$ and $SA$ is given by
$h=30$. Thus we can obtain residual entropy using eq.\eqref{eq:Sg3},
which becomes $\mathcal{S}_{c}=\ln(2)$, since $g_{1,0}^{c}=1$ and
$g_{-1,0}^{c}=1$. Another case, is the boundary between $SA$ and
$FR_{2}$ given by $h=-J_{z}-1$. The CRE at zero temperature can
be obtained using the eq.\eqref{eq:S-single-sct}, where residual
entropy becomes $\mathcal{S}_{c}=\ln(3)$, because $g_{1,0}^{c}=2$
and $g_{1,1}^{c}=1$. All the above phase boundaries are clearly discontinuous
at phase boundary \eqref{eq:S-sntd}, indicating the absence of the
pseudo-transition (see fig.\ref{fig:entropy-T0}c). In contrast, the
phase boundary between $FI$ and $FR_{2}$ is described by $J_{z}=-15$
(red dashed line). The CRE satisfying the relation \eqref{eq:Sg-psd}
becomes ${\cal S}_{c}=\ln(2)$, since the adjacent phases degeneracies
are given by $g_{1,0}^{c}=2$ and $g_{-1,0}^{c}=1$, which is in accordance
with fig.\ref{fig:entropy-T0}b.

\subsection{Thermodynamics}

Let us proceed to a discussion of the free energy \eqref{eq:free-energ}.
For the present model, the Boltzmann factors are taken using the energy
levels given by (\ref{eq:e-Trd-0}-\ref{eq:e-Trd-5}),
\begin{equation}
w_{n}=\sum_{k=0}^{5}g_{n,k}{\rm e}^{-\beta\varepsilon_{n,k}}.
\end{equation}
Hence, we have reached the following Boltzmann factors
\begin{alignat}{1}
w_{n}= & 2{\rm e}^{\beta\left(\frac{nh}{2}-\frac{J_{z}}{4}\right)}\left\{ \left({\rm e}^{\beta J}+2{\rm e}^{-\beta J/2}\right)\cosh\left(\beta\tfrac{J_{0}+h_{z}}{2}\right)\right.\nonumber \\
 & \left.+{\rm e}^{\beta J_{z}}\cosh\left(\tfrac{3}{2}\beta(J_{0}n+h_{z})\right)\right\} ,
\end{alignat}
where $n=\{-1,0,1\}$.

Therefore, we can find the system entropy at finite temperature is
$\mathcal{S}=-\frac{\partial f}{\partial T}$ . Similarly, one can
find Heisenberg spin magnetization given by $m_{H}=-\frac{\partial f}{\partial h_{z}}$
and Ising spin magnetization becomes $m_{I}=-\frac{\partial f}{\partial h}$.

In fig.\ref{fig:(a)-Zero-temperature}b we illustrate the density
plot of entropy as a function of $J_{z}$ and $h$, for fixed $T=0.6$,
and using the same scale of fig.\ref{fig:(a)-Zero-temperature}a.
Here we can observe that entropy follows the vestige of the zero temperature
phase diagram. Definitely, the thermal excitation influences the phase
boundaries, and all except one, display an increase in entropy around
the phase boundaries. The entropy at the boundary between $qFI$ and
$qFR_{2}$ (here we add the prefix \textquotedbl$q$\textquotedbl{}
to name the quasi-phases) holds almost unaltered. Because at zero
temperature, residual entropy is continuous, at least from the one-sided
limit at the phase boundary.

\begin{widetext}

\begin{figure}
\centering{}\includegraphics[scale=0.85]{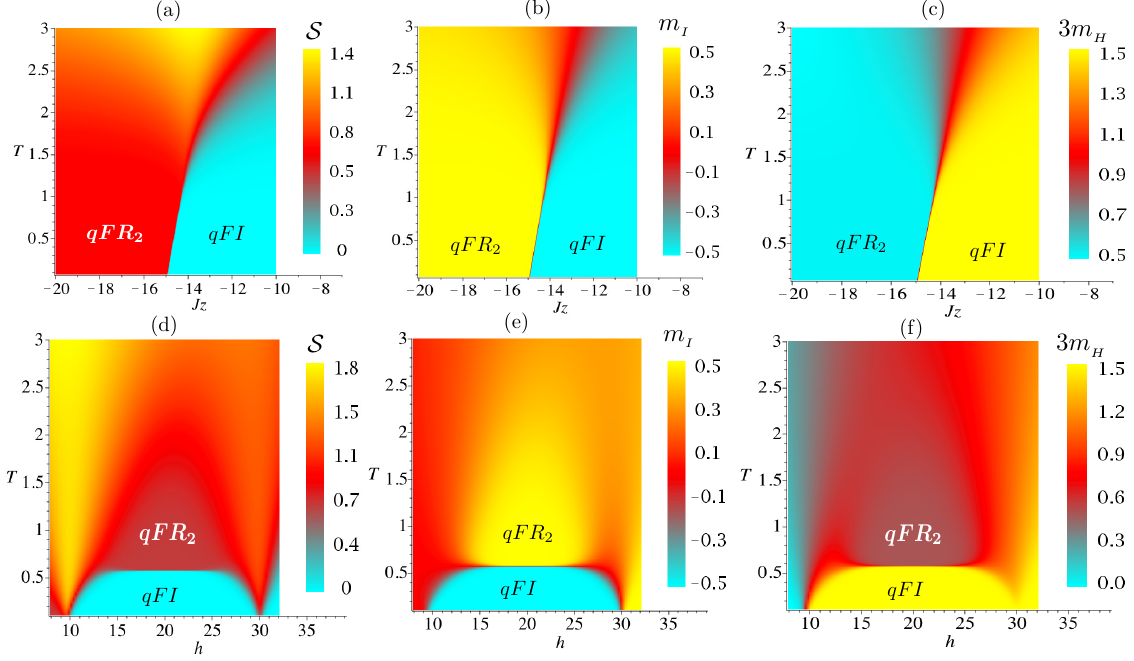}\caption{\label{fig:phs-dgm-T}Density plot for Ising-Heisenberg coupled tetrahedral
chain in the  $J_{z}-T$ plane, assuming fixed $J=-10$, $J_{0}=-10$:
(a) entropy for $h=20$; (b) Ising spin magnetization for $h=20$;
(c) Heisenberg spin magnetization for $h=20$; (d) entropy in the
$h-T$ plane for fixed $J_{z}=-14.6$; (e) Ising spin magnetization
for $J_{z}=-14.6$; (f) Heisenberg spin magnetization for $J_{z}=-14.6$.}
\end{figure}

\end{widetext}

Fig.\ref{fig:phs-dgm-T}(left column) reports density plot of entropy
in the $J_{z}-T$ plane (panel a) and $h-T$ plane (panel d), for
the parameters considered in the caption. Panel (a) for $-15<J_{z}\lesssim-14$
exhibits a pseudo-transition between quasi-phases $qFR_{2}$ and $qFI$,
for $J_{z}\gtrsim-14$ the sharp boundary melt. Panel (d) for $15\lesssim h\apprle25$,
we observe a sharp boundary between quasi-phases $qFR_{2}$ and $qFI$,
for other values of magnetic field the boundaries melt. In fig.\ref{fig:phs-dgm-T}(middle
column) is illustrated the Ising spin magnetization $m_{I}$ in the
$J_{z}-T$ plane (panel b), and in the $h-T$ plane (panel e). While
the right column reports for Heisenberg spin magnetization ($m_{H}$)
in the $J_{z}-T$ plane (panel c), and in the $h-T$ plane (panel
f). So in all panels the phase boundary is easily identified by sharp
boundaries. 

\begin{figure}
\includegraphics[scale=0.75]{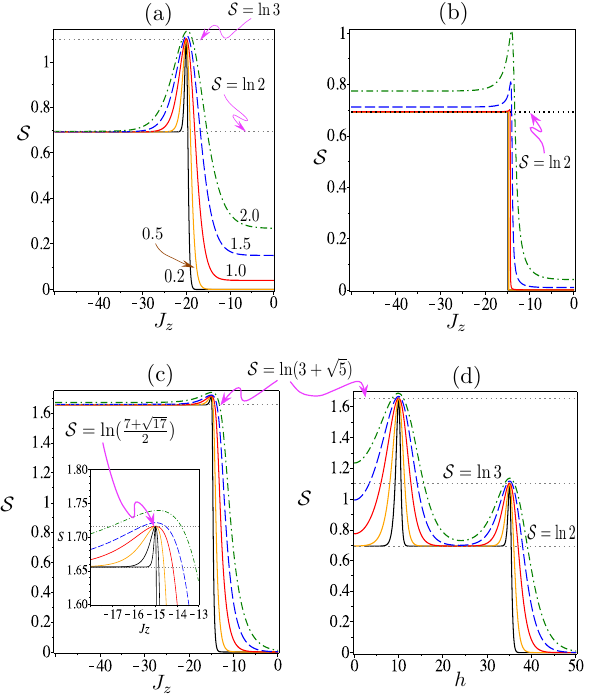}\caption{\label{fig:Entrp-tetra}(a-c) Entropy for Ising-Heisenberg coupled
tetrahedral chain as a function of $J_{z}$ assuming fixed $J=-10$,
$J_{0}=-10$, for a set of temperatures $T=\{0.2,0.5,1.0,1.5,2.0\}$:
(a) for $h=35$; (b) for $h=20$; (c) for $h=10$; (d) entropy as
a dependence of $h$ for fixed $J=-10$, $J_{0}=-10$, $J_{z}=-20$
and for the same set of temperatures in (a-c).}
\end{figure}

In fig.\ref{fig:Entrp-tetra}a is showed the entropy as a function
of $J_{z}$ in the low temperatures region. We can observe the track
of frustrated ($FR_{2}$) phase which is a macroscopically degenerate
state, whose residual entropy is $\mathcal{S}=\ln(2)$. The peak corresponds
to the phase boundary between $FR_{2}$ and $SA$, with its respective
CRE given by $\mathcal{S}_{c}=\ln(3)$. Here we can see how entropy
at finite temperature stretched around discontinuous CRE owing to
thermal excitation (see fig.\ref{fig:entropy-T0}c). In fig.\ref{fig:Entrp-tetra}b
entropy is illustrated as a dependence of $J_{z}$, where phases $FR_{2}$
and $FI$ have residual entropy $\mathcal{S}=\ln(2)$ and $\mathcal{S}=0$,
respectively. Note that entropy remains almost unaffected at $\mathcal{S}_{c}=\ln(2)$
for $T\lesssim1$ because the residual entropy at zero-temperature
is continuous from the one-sided limit at the critical point (see
fig.\ref{fig:entropy-T0}b). In fig.\ref{fig:Entrp-tetra}c, we report
entropy as a function of temperature at the interface between $FR_{1}$
{[}$\mathcal{S}=\ln(2)${]} and $FR_{2}$ {[}$\mathcal{S}=\ln(2)${]},
and $\mathcal{S}_{c}=\ln(3+\sqrt{5})$ gives the corresponding CRE
for $J_{z}\lesssim-20$. Whereas for $J_{z}=-15$, the phase boundary
joins three phases $FR_{1}$, $FR_{2}$ and $FI$, at first glance
this is similar to fig.\ref{fig:Entrp-tetra}b. But the CRE at zero
temperature is larger than the adjacent phases residual entropies,
which can be obtained using the eq.\eqref{eq:Sg3}. Since the degeneracies
of each sector are $g_{1,0}^{c}=4$, $g_{-1,0}^{c}=3$ and $g_{0,0}^{c}=2$,
under these circumstances the critical residual entropy becomes $\mathcal{S}_{c}=\ln(\frac{7+\sqrt{17}}{2})$.
We can observe better this point in a magnified plot in the inner
part of the fig.\ref{fig:Entrp-tetra}c. Because of this small peak,
the right side curve stretched around CRE for $T>0$, destroying any
evidence of pseudo-transition. At last, in fig.\ref{fig:Entrp-tetra}d
is plotted the entropy as a function of magnetic field $h$. Then
we observe a residual entropy between phase boundaries, which are
in agreement with the previous plots (panels a and c). 

In synthesis, it is worth mentioning that the entropy {[}fig.\ref{fig:Entrp-tetra}(a
and c){]} at zero temperature falls for a type of function described
in fig.\ref{fig:entropy-T0}c, while fig.\ref{fig:Entrp-tetra}d falls
for a kind of function described in fig.\ref{fig:entropy-T0}(c-d)
confirming the absence of pseudo-transition. In contrast, fig.\ref{fig:Entrp-tetra}b
fits into a type of fig.\ref{fig:entropy-T0}b, which means the existence
of pseudo-transition.

\begin{figure}
\includegraphics[scale=0.9]{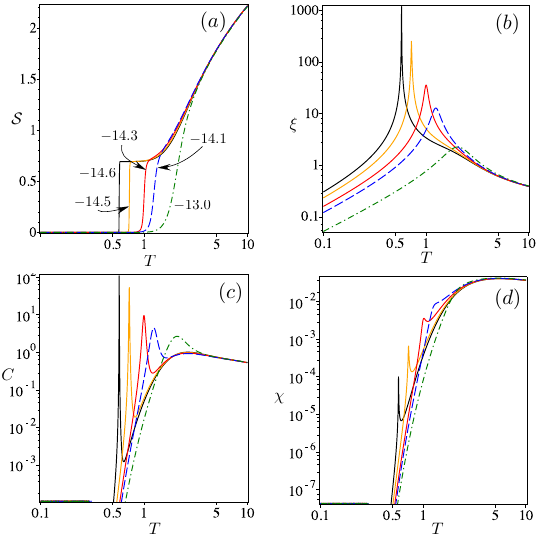}\caption{\label{fig:Phs-qnty-T}(a) Entropy ${\cal S}$ for Ising-Heisenberg
coupled tetrahedral chain as a function of temperature, assuming fixed
parameters $J=-10$, $J_{0}=-10$, $h=20$ and several values of $J_{z}=\{-13,-14.1,-14.3,-14.5,-14.6\}$
(semi-logarithmic plot); (b) correlation length $\xi$ as a dependence
of temperature (logarithmic plot); (c) specific heat $C$ against
temperature (logarithmic plot); (d) magnetic susceptibility $\chi$
as a function of temperature (logarithmic plot). }
\end{figure}

Fig.\ref{fig:Phs-qnty-T}a depicts entropy as a function of temperature
assuming fixed parameters $J=-10$, $J_{0}=-10$, $h=20$ and for
several values $J_{z}=\{-13,-14.1,-14.3,-14.5,-14.6\}$. It is evident
to observe a strong change in the entropy curvature at the pseudo-critical
temperature. Thus, as the temperature rises, the sharp jump in entropy
becomes softer and gradually vanishes. In fig.\ref{fig:Phs-qnty-T}b
is illustrated the correlation length as a temperature dependence,
so that it corroborates a sharp and robust peak at pseudo-critical
temperature. In fig.\ref{fig:Phs-qnty-T}c is reported the specific
heat as a function of temperature, and once again, we observe a sharp
peak at pseudo-critical temperature. Whereas in fig.\ref{fig:Phs-qnty-T}d
is reported the magnetic susceptibility as a function of temperature,
which exhibits a small sharp peak. Since the total magnetization at
zero temperature is identical for both phases $FR_{2}$ and $FI$
according to \eqref{Trt-FI} and \eqref{Trt-FR2}, respectively.

\section{Ising-XYZ diamond chain}

\begin{figure}[h]
\centering{}\includegraphics[scale=0.8]{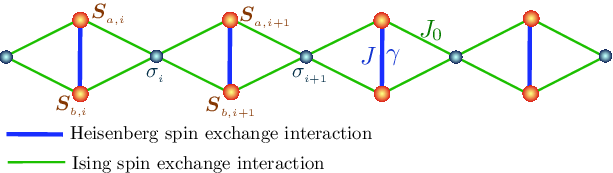}\caption{\label{schm-dmd}Schematic representation of Ising-XYZ diamond chain.
Small spheres ($\sigma_{i}$) correspond to Ising spins and large
spheres ($\boldsymbol{S}_{a(b),i}$) correspond to Heisenberg spins.}
\end{figure}

Another model we consider here is the Ising-XYZ diamond chain structure,
as illustrated in fig.\ref{schm-dmd}, which was discussed earlier
in reference \citep{pseudo,torrico,torrico2}. Here $\sigma_{i}$
(small spheres) represents the Ising spin-1/2, and $S_{a(b),i}^{\alpha}$
(large spheres) denotes the Heisenberg spin-1/2, with $\alpha=\{x,y,z\}$.
Despite Ising-XXZ diamond chain is frustrated owing to triangular
structure and $xy$-isotropy, Ising-XYZ diamond chain also exhibits
unfrustrated phase boundary\citep{pseudo,torrico,torrico2}, because
of $xy$-anisotropy. Thus, here we only give a revisiting of Ising-XYZ
diamond chain, whose Hamiltonian is expressed by 
\begin{alignat}{1}
H=-\sum_{i=1}^{N} & \left[J(1+\gamma)S_{a,i}^{x}S_{b,i}^{x}+J(1-\gamma)S_{a,i}^{y}S_{b,i}^{y}\right.\nonumber \\
 & +J_{z}S_{a,i}^{z}S_{b,i}^{z}+J_{0}(S_{a,i}^{z}+S_{b,i}^{z})(\sigma_{i}+\sigma_{i+1})\nonumber \\
 & \left.+h_{z}(S_{a,i}^{z}+S_{b,i}^{z})+\tfrac{h}{2}(\sigma_{i}+\sigma_{i+1})\right],\label{eq:Hamt}
\end{alignat}
where $J$ corresponds to $xy$-axis exchange interaction and $\gamma$
being the $xy$-anisotropy, $J_{z}$ stands for Heisenberg spins exchange
interaction on the $z$-axis. While $J_{0}$ denotes Ising-Heisenberg
spin exchange interaction, and $h_{z}$($h$) corresponds to the external
magnetic field along the $z$-axis acting on Heisenberg spin (Ising
spin), respectively. 

\subsection{Zero temperature phase diagram}

Further investigation of the ground state phase diagram has already
been found in reference \citep{torrico,torrico2}. Below are summarized
some of those ground states assuming $n=\sigma_{i}+\sigma_{i+1}$:

(i) For sector $n=1$ ($\uparrow\uparrow$ ) the first ground state
energy is 
\begin{equation}
\varepsilon_{1,0}=E_{_{MF_{2}}}=-\tfrac{J_{z}}{4}-\tfrac{h}{2}-\sqrt{(h_{z}+J_{0})^{2}+\tfrac{1}{4}J^{2}\gamma^{2}}.
\end{equation}
Named as modulated ferromagnetic Heisenberg spin ($MF_{2}$) phase.
With its corresponding ground state 
\begin{alignat}{1}
\left|MF_{2}\right\rangle = & \overset{N}{\underset{i=1}{\prod}}\left(\cos\theta_{1}|\begin{smallmatrix}+\\
+
\end{smallmatrix}\rangle_{i}+\sin\theta_{1}|\begin{smallmatrix}-\\
-
\end{smallmatrix}\rangle_{i}\right)\otimes|\uparrow\rangle_{i},\label{eq:state3}
\end{alignat}
where $\theta_{n}=\frac{1}{2}\tan^{-1}\frac{J\gamma}{2(h_{z}+J_{0}n)}$
defined in $-\frac{\pi}{4}<\theta_{n}<\frac{\pi}{4}$. 

Another state is the ferrimagnetic (FI) phase, which is given by
\begin{equation}
\varepsilon_{1,1}=E_{_{FI}}=-\tfrac{J+h}{2}+\tfrac{J_{z}}{4},
\end{equation}
and the corresponding ground state is expressed as 
\begin{alignat}{1}
\left|FI\right\rangle = & \overset{N}{\underset{i=1}{\prod}}\frac{1}{\sqrt{2}}\left(|\begin{smallmatrix}-\\
+
\end{smallmatrix}\rangle_{i}+|\begin{smallmatrix}+\\
-
\end{smallmatrix}\rangle_{i}\right)\otimes|\uparrow\rangle_{i}.\label{eq:state4}
\end{alignat}

(ii) For sector $n=-1$ ($\downarrow\downarrow$), the ground state
energy, becomes
\begin{equation}
\varepsilon_{-1,0}=E_{_{MF_{0}}}=-\tfrac{J_{z}}{4}+\tfrac{h}{2}-\sqrt{(h_{z}-J_{0})^{2}+\tfrac{1}{4}J^{2}\gamma^{2}},
\end{equation}
with its respective modulated ferromagnetic ($MF_{0}$) state, given
by 
\begin{alignat}{1}
\left|MF_{0}\right\rangle = & \overset{N}{\underset{i=1}{\prod}}\left(\cos\theta_{-1}|\begin{smallmatrix}+\\
+
\end{smallmatrix}\rangle_{i}+\sin\theta_{-1}|\begin{smallmatrix}-\\
-
\end{smallmatrix}\rangle_{i}\right)\otimes|\downarrow\rangle_{i}.\label{eq:state2}
\end{alignat}
Now assuming $h_{z}=h$, the ground state energy at the interface
between $MF_{0}$ and $MF_{2}$ must coincide (see fig.\ref{fig:Phase-diagram}),
implying that there is a critical magnetic field given by

\begin{equation}
h_{c}=\frac{(\gamma^{2}-1)+2J_{z}J+4J_{0}^{2}-J_{z}^{2}}{4J+8J_{0}-4J_{z}}.
\end{equation}
 For $h<h_{c}$ the system is in $MF_{0}$ state, and for $h>h_{c}$
the system becomes in $MF_{2}$ state. 

\subsection{Thermodynamics}

Next, let us take a look at the thermodynamics of the model. Thus,
the free energy \eqref{eq:free-energ} for the Ising-XYZ diamond chain
was also obtained in reference \citep{torrico,torrico2}, where the
Boltzmann factors were given by 
\begin{equation}
w_{n}=2{\rm e}^{\frac{\beta nh}{2}}\left[{\rm e}^{-\frac{\beta J_{z}}{4}}{\rm ch}\left(\tfrac{\beta J}{2}\right)+{\rm e}^{\frac{\beta J_{z}}{4}}{\rm ch}\left(\beta\Delta_{n}\right)\right],
\end{equation}
with $\Delta_{n}=\sqrt{(h_{z}+J_{0}n)^{2}+\frac{1}{4}J^{2}\gamma^{2}}$. 

\begin{widetext}

\begin{figure}
\centering{}\includegraphics[scale=0.8]{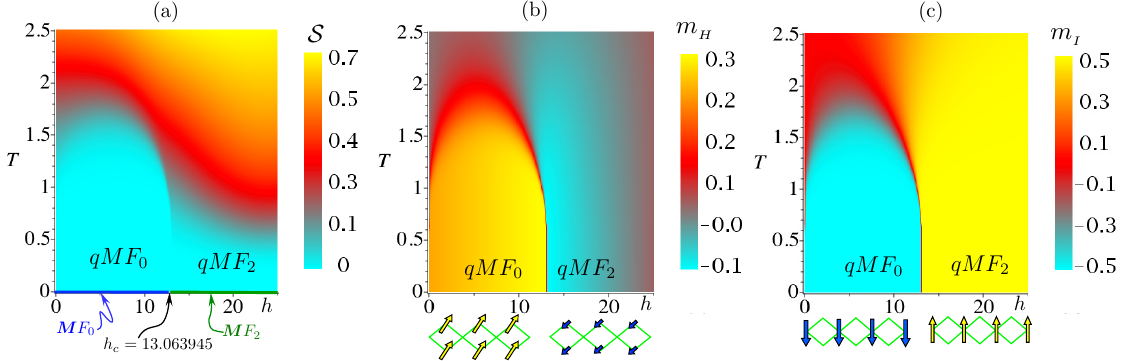}\caption{\label{fig:Phase-diagram}Density plot for Ising-XYZ diamond chain
assuming fixed parameters $J=100$, $\gamma=0.8$, $J_{z}=24$ and
$J_{0}=-24$ in the $h-T$ plane: (a) entropy; (b) Heisenberg spin
magnetization; (c) Ising spin Magnetization.}
\end{figure}

\end{widetext}

In fig.\ref{fig:Phase-diagram}a we illustrate the density plot entropy
as a function of temperature for fixed parameters $J=100$, $\gamma=0.8$,
$J_{z}=24$ and $J_{0}=-24$. The density plot of entropy in low-temperature
region shows an indistinguishable boundary between quasi-phases $qMF_{0}$
and $qMF_{2}$. Because, at zero temperature, the ground state energy
for both phases $MF_{0}$ and $MF_{2}$ are non-degenerate, which
implies that the residual entropy is ${\cal S}=0$. Besides, CRE also
becomes null according to relation \eqref{eq:Sg-psd}, which is easily
verified on the density plot of entropy.

However, the density plot of Heisenberg spin magnetization ($m_{H}$)
in the plane $h-T$ is reported in fig.\ref{fig:Phase-diagram}b.
Doubtless, the boundary between quasi-phases $qMF_{2}$ and $qFM_{0}$
shows a distinguishable region.  At $T=0$ and $h<h_{c}=13.063045$,
the Heisenberg spins are parallel ordered with greater probability
pointing up. The maximum magnetization per spin is $m_{H}\sim0.3$,
which is pictorially denoted as effective canting spin (yellow region).
Similarly, for $h>h_{c}=13.063945$ the Heisenberg spin magnetization
becomes negative $m_{H}\sim-0.1$, pictorially illustrated by effective
canting spin down (cyan region). For more detailed information concerning
Heisenberg spin magnetization, we refer the readers to ref. \citep{Isaac}.
In a similar way the fig.\ref{fig:Phase-diagram}c displays the density
plot of Ising spin magnetization, for $h<h_{c}$ the magnetization
is nearly $m_{I}=-0.5$, which means most of the Ising spins are aligned
downward. While for $h>h_{c}$ most Ising spins are pointing up, aligning
with the external magnetic field. Consequently, this behavior could be
 easily misinterpreted as a true phase transition. Although, we do
not expect a genuine phase transition at finite temperature, because
all free energy derivatives are analytical. It is noteworthy that,
at finite temperature, there is no critical magnetic field. However,
only a pseudo-critical magnetic field $h_{p}\lesssim h_{c}$, which
vanishes roughly around $T\sim1.0$, for temperature $T\gtrsim1.0$,
the system becomes a standard disordered system predominantly. 

\begin{figure}
\includegraphics[scale=0.79]{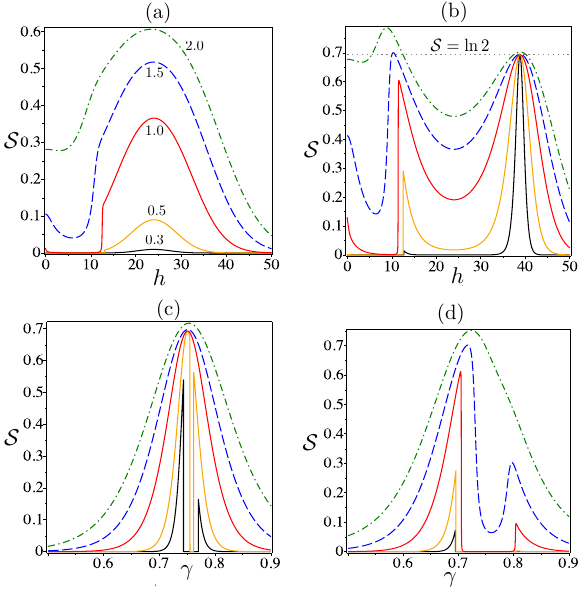}\caption{\label{fig:Ent-h-g}Entropy for Ising-XYZ diamond chain considering
fixed parameters $J=100$, $J_{z}=24$, $J_{0}=-24$ and several values
of temperatures $T=\{0.3,0.5,1.0,1.5,2.0\}$: (a) as a function of
$h$ considering $\gamma=0.8$; (b) as a function of $h$ assuming
$\gamma=0.7$; (c) as a function of $\gamma$ considering $h=18$;
(d) as a function of $\gamma$ for a fixed $h=12$.}
\end{figure}

Fig.\ref{fig:Ent-h-g}a illustrates the entropy as a function of $h$,
for a set of temperature values $T=\{0.3,0.5,1.0,1.5,2.0\}$. It can
be seen from the curve, that there is a sudden change for $T\lesssim1.0$,
this one corresponds to pseudo-transition at $h=13.063945$. Therefore,
for any region or quasi-phase, the entropy vanishes when $T\rightarrow0$.
Similarly, in fig.\ref{fig:Ent-h-g}b entropy is plotted as a dependence
of $h$, where we observe a continuous sudden change to the magnetic
field $h=12.8$. Again, as soon as temperature decreases, the entropy
vanishes according to fig.\ref{fig:entropy-T0}a. However, for $h\approx39$
corresponds to the phase boundary between $FI$ and $MF_{2}$\citep{pseudo,torrico,torrico2}
with a critical residual entropy $\mathcal{S}_{c}=\ln(2)$, and obviously
in this boundary there is no pseudo-transition (see fig.\ref{fig:entropy-T0}d).
Furthermore, in fig.\ref{fig:Ent-h-g}c-d, entropy against $\gamma$
is plotted for $h=18$ and $h=12$. For temperature $T\lesssim1$,
a continuous sudden change appears, showing the pseudo-transition.
Entropy vanishes when $T\rightarrow0$, because the CRE becomes null,
which is in accordance with fig.\ref{fig:entropy-T0}a. In reference
\citep{Isaac,pseudo,torrico,torrico2}, the reader can find other
detailed discussions concerning this model. 

\section{Conclusions.}

Although few one-dimensional models exhibit the phase transition\citep{kittel,chui,dauxois,sarkanych},
this phenomenon is related to some null elements of the transfer-matrix,
which leads to a non-analytic free energy. Nevertheless, pseudo-critical
temperatures have recently been investigated in one-dimensional spin
models\citep{Isaac,pseudo}. Therefore, there are several models exhibiting
pseudo-transitions, such as Ising-Heisenberg spin models with a variety
of structures\citep{torrico,torrico2,Galisova,on-strk,strk-cav}. 

We propose here a relationship between critical residual entropy at
zero temperature and pseudo-transition. In general, residual entropy
increases at the interface where the phase transition occurs at a
critical point. Which means the system increases its ground state
degeneracy in the interface compared to adjacent states. In contrast,
there are few cases where CRE remains equal to the largest residual
entropy of neighboring states, here CRE is given by $\mathcal{S}_{c}=\ln\left[\max\left(g_{1,0},g_{-1,0}\right)\right]$.
So our main result dwells in a simple argument to recognize a pseudo-transition.
If the zero temperature residual entropy is continuous at least from
the one-sided limit at critical point, then analytical free energy
keep in sight the evidences of a pseudo-transition at finite temperature
around critical point. In order to show the aforementioned property,
we considered two Ising-Heisenberg spin models: one frustrated model
in a coupled tetrahedral chain and another unfrustrated diamond chain. 

Finding pseudo-transition in more realistic systems, like the quantum
Heisenberg model, would be a fascinating investigation. Nevertheless,
this would be a cumbersome numerical task at finite temperature. However,
searching for residual entropy at zero temperature should be an easier
task than studying the complete thermodynamics of the model. Only
after this analysis would it be possible to study thermodynamics for
a particular condition previously analyzed at zero temperature. In
this sense, it would be interesting whether the condition of the phase
boundary entropy still holds. Assuming our argument is still valid,
we can apply this condition at zero temperature and look for continuity
of residual entropy, which would be a more manageable task than studying
the complete thermodynamics of the model.
\begin{acknowledgments}
This work was partially supported by ICTP and Brazilian agencies CNPq
and FAPEMIG.
\end{acknowledgments}

\appendix

\section{\label{Appdx-A}Discontinuous residual entropy}

When CRE is larger than its neighboring residual entropy at zero temperature,
it means a discontinuous residual entropy because of $G_{c}>\max(g_{1,0},g_{-1,0})$.
Then, to obtain the residual entropy we can use the free energy \eqref{eq:free-energ}
in the limit $T\rightarrow0$. Below we get for some particular cases,
the residual entropy at zero temperature.

\subsection{Phase boundary between states of sector $n=0$ and $n=\pm1$}

Here let us consider the sector $n=1$ and $n=0$, where the lowest
energies are given by $\varepsilon_{_{1,0}}(x)$ and $\varepsilon_{_{0,0}}(x)$.
Then the phase boundary occurs when $\varepsilon_{_{1,0}}(x_{c})=\varepsilon_{_{0,0}}(x_{c})=\varepsilon_{c}$
with corresponding degeneracies $g_{1,0}^{c}$ and $g_{-1,0}^{c}$.
Eventually, we could have $g_{1,0}^{c}\geqslant g_{1,0}$ and $g_{0,0}^{c}\geqslant g_{0,0}$.
The lowest energy $\varepsilon_{-1,0}(x)$ in sector $n=-1$, will
be strictly higher than $\varepsilon_{c}$ ($\varepsilon_{-1,0}>\varepsilon_{c}$),
what means $w_{-1}/w_{0}\rightarrow0$ and $w_{-1}/w_{1}\rightarrow0$
when $T\rightarrow0$. So the free energy in \eqref{eq:free-energ}
at sufficiently low temperature becomes
\begin{equation}
f=-\tfrac{1}{\beta}\ln\left[\tfrac{1}{2}\Bigl(g_{1,0}^{c}+\sqrt{(g_{1,0}^{c})^{2}+4(g_{0,0}^{c})^{2}}\Bigr){\rm e}^{-\beta\varepsilon_{c}}\right].
\end{equation}
Consequently, the CRE turns in 
\begin{equation}
\mathcal{S}_{c}=\ln\left[\tfrac{1}{2}\Bigl(g_{1,0}^{c}+\sqrt{(g_{1,0}^{c})^{2}+4(g_{0,0}^{c})^{2}}\Bigr)\right],\label{eq:S_01}
\end{equation}
 where the critical degeneracy results in $G_{c}=\tfrac{1}{2}\Bigl(g_{1,0}^{c}+\sqrt{(g_{1,0}^{c})^{2}+4(g_{0,0}^{c})^{2}}\Bigr)$.

We can observe the critical degeneracy is strictly larger than its
adjacent degeneracies: $G_{c}>g_{1,0}$ and $G_{c}>g_{0,0}$. The
residual entropy is reported schematically in fig.\ref{fig:entropy-T0}c.
Note that the, residual entropy ${\cal S}_{a}$ and ${\cal S}_{b}$
denote in general a residual entropy between adjacent states. 

For sector $n=-1$ and $n=0$, the result of free energy will be equivalent
to the previous case. Therefore we can obtain merely by exchanging
$\varepsilon_{_{1,0}}(x_{c})\rightarrow\varepsilon_{_{-1,0}}(x_{c})$,
in expression \eqref{eq:S_01}.

\subsection{Phase boundary lying in a single sector}

Assuming that occurs a phase transition between two states with energies
$\varepsilon_{1,0}(x)$ and $\varepsilon_{1,1}(x)$ , then the critical
energy is given by $\varepsilon_{1,0}(x_{c})=\varepsilon_{1,1}(x_{c})=\varepsilon_{c}$
at $T=0$. In general, it is possible that some additional states
can coincide at the phase boundary, then the degeneracies can be eventually
expressed satisfying the condition $g_{1,0}^{c}\geqslant g_{1,0}$
and $g_{1,1}^{c}\geqslant g_{1,1}$. Therefore, all other energy levels
must be higher than $\varepsilon_{c}$, so when $T\rightarrow0$,
the spectral energy in other sectors can be neglected ($w_{0}/w_{1}\rightarrow0$
and $w_{-1}/w_{1}\rightarrow0$). Hence, the free energy in the low
temperature limit is expressed as 
\begin{alignat}{1}
f= & -\tfrac{1}{\beta}\ln\left(w_{1}\right)=-\tfrac{1}{\beta}\ln\left(g_{1,0}^{c}{\rm e}^{-\beta\varepsilon_{c}}+g_{1,1}^{c}{\rm e}^{-\beta\varepsilon_{c}}\right)\nonumber \\
= & -\tfrac{1}{\beta}\ln\left[\left(g_{1,0}^{c}+g_{1,1}^{c}\right){\rm e}^{-\beta\varepsilon_{c}}\right].
\end{alignat}
Whereas, the corresponding critical residual entropy, reduces to
\begin{equation}
\mathcal{S}_{c}=\ln\left(g_{1,0}^{c}+g_{1,1}^{c}\right).\label{eq:S-single-sct}
\end{equation}
Thereby, the critical degeneracy is given by $G_{c}=(g_{1,0}^{c}+g_{1,1}^{c})$.

Once again, the CRE is strictly higher than any residual entropy of
adjacent states, because $G_{c}>g_{1,0}^{c}$ and $G_{c}>g_{1,1}^{c}$.
A schematic representation of this type of CRE is illustrated in fig.\ref{fig:entropy-T0}c.

\subsection{Phase boundary lying in three sectors}

When three sectors can constitute the phase boundary, the states with
energies $\varepsilon_{_{1,0}}(x)$, $\varepsilon_{_{0,0}}(x)$ and
$\varepsilon_{_{-1,0}}(x)$, can coexist for a particular $x_{c}$
(a critical point). So we must assume $\varepsilon_{_{1,0}}(x_{c})=\varepsilon_{_{0,0}}(x_{c})=\varepsilon_{_{-1,0}}(x_{c})=\varepsilon_{c}$,
and the respective degeneracies are $g_{1,0}^{c}$, $g_{0,0}^{c}$
and $g_{-1,0}^{c}$. In this case, the free energy becomes 
\begin{alignat}{1}
f= & -\tfrac{1}{\beta}\ln\left[\tfrac{1}{2}\Bigl(g_{1,0}^{c}+g_{-1,0}^{c}+\right.\nonumber \\
 & \left.+\sqrt{(g_{1,0}^{c}-g_{-1,0}^{c})^{2}+4(g_{0,0}^{c})^{2}}\Bigr){\rm e}^{-\beta\varepsilon_{c}}\right].
\end{alignat}
Finally, the CRE reads 
\begin{equation}
\mathcal{S}_{c}=\ln\left[\tfrac{1}{2}\Bigl(g_{1,0}^{c}+g_{-1,0}^{c}+\sqrt{(g_{1,0}^{c}-g_{-1,0}^{c})^{2}+4(g_{0,0}^{c})^{2}}\Bigr)\right].\label{eq:Sg3}
\end{equation}
Then the critical degeneracy is $G_{c}=\tfrac{1}{2}\Bigl(g_{1,0}^{c}+g_{-1,0}^{c}+\sqrt{(g_{1,0}^{c}-g_{-1,0}^{c})^{2}+4(g_{0,0}^{c})^{2}}\Bigr)$.

Similar to the previous cases, the CRE is strictly larger than its
adjacent states residual entropies, because $G_{c}>g_{1,0}^{c}$,
$G_{c}>g_{0,0}^{c}$ and $G_{c}>g_{-1,0}^{c}$. 

In all aforementioned cases, the CRE inevitably exhibits a jump-point
discontinuity or a point discontinuity at $x_{c}$ (see fig.\ref{fig:entropy-T0}c-d).
Because the CRE is strictly larger than its adjacent states residual
entropy.

\vspace{1.cm}

\subsection*{\hskip-1.5cmReference}

\end{document}